\documentclass[12pt]{article}
\usepackage{amssymb}
\usepackage{amsmath}
\usepackage{hyperref}
\begin{document}
\title{\bf Magnetic charge quantization from SYM considerations}
\author{Milad Porforough\thanks{Email: milaad@aut.ac.ir} \hspace{1mm}\\
		{\small {\em  Department of Physics, School of Sciences,}}\\
        {\small {\em Tarbiat Modares University, P.O.Box 14155-4838, Tehran, Iran}}\\
       }
\date{\today}
\maketitle

\abstract{An intersecting D3-D3' system contains magnetic monopole solutions due to D- strings stretched between two branes. These magnetic charges satisfy the usual Dirac quantization relation. We show that this quantization condition can also be obtained directly by SUSY and gauge invariance arguments of the theory and conclude that the independence of physics from a shift of holonomy is exactly equivalent to regarding a {\it Fayet-Iliopoulos (FI) gauge} for our set-up. So we are led to conjecture that there is a correspondence between the topological point of view of magnetic charges and SYM considerations of their theories.  This picture implies that one can attribute a definite quantity to the integration of the vector multiplet over the singular region such that we can identify it with magnetic flux.  It also indicates that the FI parameter is proportional to the magnetic charge so it is a quantized number.\\

\section{Introduction} \label{intro}
The Hanany-Witten brane construction \cite{Ref1} gives an explanation for the relation between three-dimensional gauge theories with $\mathcal{N}=4$ supersymmetry and the moduli space of $n$, BPS $SU(2)$ monopoles. There are $N$, D3-branes stretched between two parallel NS5-branes. NS5-branes are extended in (012345) directions while D3-branes are along (0126) directions.

Infinite directions for each D3-brane are (01) and (6), so the macroscopic field theory for this is $\mathcal{N}= 4, 2+1$-$d$, $U(N )$ gauge theory.  Location of a D3-brane is specified by $x = (x^3, x^4, x^5)$ values which can be regarded as the expectation values of the three scalar fields of a vector multiplet in adjoint representation. In the Coulomb branch of a gauge theory where $U(N)$ gauge symmetry is broken to $U(1)^N$, each of these N photons corresponds to a periodic scalar so the Coulomb branch is a $4N$-dimensional space. The low-energy effective dynamics of the gauge theory is completely determined by the metric of this space.

One can determine the metric of the Coulomb branch by having different perspectives on the brane picture. We can start by  performing a S-duality (or weak-strong duality) such that the NS5-branes become D5-branes. The worldvolume theory of D5-branes is a six dimensional $SU(2)$ Yang-Mills theory. There are $N$ magnetic monopoles due to D3-branes stretched between them \cite{Ref2}. It proposes that the metric on the Coulomb branch of the 3d, $SU(N)$ SYM theory, which is corrected by quantum considerations, corresponds to the metric of the classical moduli space of those N monopoles.

We can change the brane set-up such that the worldvolume of the D3-brane becomes a 3d SYM theory with $\mathcal{N}=2$ supersymmetry. For this purpose, we should rotate just one of the NS5-branes \cite{Ref3,Ref4,Ref5}, to reach another NS5 which is usually referred to as NS5'-brane such that its worldvolume is extended over (012378) directions. The D3-branes can move only in $x^3$-direction. In a typical point on the classical moduli space, one may again dualize the $N$ photons such that there leaves $2N$ low energy dynamics of these modes.

This theory has been studied recently \cite{Ref6,Ref7,Ref8}, and we know that this system has magnetic solutions but the trouble is that this configuration does not admit any soliton solutions in the canonical description which can be identified with the D-strings stretched between D3-branes. However, the set-up was studied by Mintun, Polchinski and Sun in \cite{Ref6}, where they argued that by considering the periodicity in the hypermultiplet space, we are led to a non-trivial Gibbons-Hawking metric in the non-canonical description such that one can find the
expected magnetic kink solution.

In this paper, we will propose an alternative way to obtain Dirac quantization condition for magnetic monopoles by using SUSY and gauge invariance of the field theory of this system. In the next section, we will review the main topics of \cite{Ref6}, to understand how one can construct  a field theory for such a configuration. In the third section, we will introduce the Fayet-Iliopoulos gauge invariance for this set-up and show that it yields the Dirac quantization condition. In the fourth section, we will give  a short analysis without looking at the role of the action. The paper closes in the fifth section, with a brief concluding remark.

\section{The field theory of D3-D3' system}
\label{sec2}

In this section, in order to fix our notation, we follow \cite{Ref6}, to see how one can construct the field theory of intersecting D3-D3' branes.  Consider a D3-brane spanning the (0145) directions and an orthogonal D3'-brane spanning the (0167) directions such that eight supercharges are preserved. On each D3-brane, there lives the usual field content for a $U(1)$, $\mathcal{N}= 4, d = 4$ gauge theory, but the supersymmetry algebras of two branes are not the same.

If one tends to use the strategy introduced in  \cite{Ref9} and  \cite{Ref10} to write the full action of the theory,  he or she must first T-dualise the system in (23) directions which are orthogonal to both branes so they correspond to DD boundary conditions. Then the D3-D3' becomes a D5-D5'. Note that after writing the full action, one should dimensionally reduce it in the (23) directions to come back to the system of interest. The  T-dual  configuration  has  different  global  symmetries,  but the fact that the dimensionally reduced system will have a $SO(4)_{2389}$ symmetry, guarantees  that it has $\mathcal{N}=2, d = 4$ SUSY. So all we need to construct a  6d SYM theory for the D3 and D3'-branes in terms of $\mathcal{N} =1, d=4$ multiplets are: a vector multiplet $V$, and three chiral multiplets $Q_{1,2,3}$, for D3-brane and another vector multiplet $V'$, and three chiral multiplets $S_{1,2,3}$, which live on the D3' worldvolume. The scalars $A_{V2,3}$ and $A_{V'2,3}$, combine with the scalars $Q_3$ and $S_3$ respectively to form $SO(4)_{2389}$ vectors since these fields will describe the transverse coordinates of the branes in the (2389) directions. According to  \cite{Ref9} and  \cite{Ref10}, and after dimensional reduction in the $x^{2,3}$-directions, only the integrations over the (0145) directions remain and all fields become functions of the parameters $x_{0,1,4,5}$. $Q_1$($S_2$), will mix to $A_{V0,1}$($A_{V'0,1}$) to produce the corresponding gauge fields neatly.

Ultimately, the action for the D3-brane takes the form
\begin{align} \label{1}
S_{3\text{-}3'}=\frac{1}{g^2_{YM}} \int \text{d}^2x \; \text{d}x_4 \text{d}x_5 \Bigg[ \int \text{d}^2\theta \Big( \frac{1}{4} W^\alpha _V W_{V\alpha}+\frac{1}{2} (Q_3 \partial_{z_1} Q_2- 2\leftrightarrow 3) \Big) +c.c. \\ \nonumber
+\int \text{d}^4\theta \Big( (\sqrt{2}\bar{\partial}_{z_1}V-\bar{Q}_1) (\sqrt{2} \partial_{z_1}V-Q_1)- \bar{\partial}_{z_1}V \;\partial_{z_1}V+\bar{Q}_2 Q_2+\bar{Q}_3 Q_3 \Big) \Bigg],
\end{align}
where all $\mathcal{N}=1$, $d = 4$, chiral and vector multiplets are in the usual form as in \cite{Ref11}:
\begin{align}\label{2}
& V=-\theta \sigma_\mu \bar{\theta} A^\mu_V+i \theta^2  \; \bar{\theta}\bar{\lambda}_V-i \bar{\theta}^2 \; \theta \lambda_V+\frac{1}{2} \theta^2 \bar{\theta}^2 D_V, \\ \nonumber
&Q_3=Q_3(y)+\sqrt{2} \theta \psi_{Q_3}(y)+\theta^2 F_{Q_3}(y),
\end{align}
where for all chiral multiplets we use the same symbol for the scalar components as for the superfields themselves. In action (\ref{1}), we have $z_1=\frac{1}{2}(x_4 + ix_5)$, and $Q_1=(iA_{V4}+A_{V5})/\sqrt{2} \equiv iA_{Vz_1}/\sqrt{2}$  (see footnote \footnote {In complex coordinates, we have $\partial_{z_1}=\partial_4-i\partial_5$, so the gauge transformation becomes $A_{Vz_1} \rightarrow A_{Vz_1}+\partial_{z_1}\lambda$.}). Finally, Greek indices run over $\mu= 0,1,2,3$, and spinor dotted and undotted indices take two values as usual in 4 dimensions. One can show that this action is invariant under these gauge transformations:
\begin{align}\label{3}
& V \rightarrow V+\Lambda + \bar{\Lambda},\\ \nonumber
&Q_1 \rightarrow Q_1+\sqrt{2} \partial_{z_1}\Lambda,
\end{align}
where $\Lambda$ is a chiral gauge parameter which defines the gauge transformation as usual. The whole argument holds also for the second brane worldvolume theory.

The best part of the story, is the hypermultiplet action $S_{3-3'}$.  The simplest choice for the hypermultiplet kinetic terms takes the canonical form. In this case, the hypermultiplet consists of two fields $B$ and $C$, which live on the defect and have charges $(1,-1)$ and $(-1,1)$ under $U_V(1)\times U_{V'}(1)$, respectively, such that the action becomes
\begin{align}\label{4}
S_{3\text{-}3'}=\frac{1}{g^2_{YM}} \int \text{d}^4x \bigg[ \int \text{d}^4 \theta \big(|B|^2 e^{V-V'}+|C|^2 e^{V'-V} \big) \\ \nonumber
+\frac{i}{\sqrt{2}} \int \text{d}^2 \theta (BCQ_3-BCS_3)+c.c. \bigg].
\end{align}

As explained in \cite{Ref6}, there is a problem with the action (\ref{4}) : when D3-branes are separated, it does not admit any soliton solutions because the potential has not any non-trivial vacuum and takes its minimum value only when $B=C=0$. So the authors suggest a non-canonical action for the kinetic terms. There are some conditions that the kinetic terms in $S_{3-3'}$, should satisfy.  First, to have 8 supercharges, the target space should admit a hyper-K\"ahler metric. Second, as we know, we must couple the hypermultiplet to a $U (1)$ gauge field. This means that the metric must have a tri-holomorphic (or $U(1)$) isometry. Finally, the metric should admit an extra $U(1)_\mathcal{R}$ isometry which leads one of the three complex structures to become invariant and rotates the remainder. This guarantees that there is a $U(1)_\mathcal{R}$ $R$-symmetry in the field theory, a property which one can regard as the $U (1)_{45}\times U(1)_{67}$, rotational symmetry of the brane configuration.

\section{Fayet-Iliopoulos gauge and Dirac monopole}
\label{sec3}
Now we are in the position to define the Fayet-Iliopoulos gauge: once the gauge group of a SYM theory is $U(1)_1 \times \cdots \times U (1)_n$, one allows to add $V^1+\cdots+V^n$, to the D-terms of the action where $V^i$'s denote the vector multiplets in the Abelian case.  These are Fayet-Iliopoulos terms \cite{Ref12}.  Under an Abelian gauge transformation $V^i\rightarrow V^i+\Lambda + \bar{\Lambda}$,  and by keeping in mind that for a chiral superfield $\Lambda$, the only thing survived in the D-term integration is $-\frac{1}{4}\theta^2 \bar{\theta}^2 \partial^2 \Lambda$, it is clear that the FI Lagrangian
\begin{equation}\label{5}
\mathcal{L}_{FI}=\sum_{i\in abelian factors} \chi^i \int \text{d}^4 \theta \; V^i=\frac{1}{2} \sum_{i\in abelian factors} \chi^i D_{V^i},
\end{equation}
is SYM invariant and in this case $\chi$'s are constant functions and $D_{V^i}$ is the auxiliary field in the corresponding vector multiplet $V^i$ of the form (\ref{2}), as usual. Now let us come back to our set-up. Since the gauge group of the SYM theory on the defect is $U_V(1) \times U_{V'}(1)$, we can add  $(\chi_V V + \chi_{V'} V')$ to the first term in (\ref{4}), where $\chi$'s live on
the defect \footnote {According to (\ref{4}), we have defined $\chi$'s, up to a factor $g^2_{YM}$. As we will see, we work with a 2d dimensionless FI-term, because of a dimensional reduction. If we need to express our results in terms of the usual 4d FI-term, we should add a factor $M^2_{Planck}$.}.

For the D3-brane, the term which includes the total derivative is
\begin{equation}\label{6}
\frac{1}{2} \int \text{d}^2x  \;\chi_V \; \partial_\mu \partial^\mu z,
\end{equation}
where $z$ is the complex scalar field in the chiral gauge parameter $\Lambda$. It is convenient to decompose the partial components into normal $n=2,3$ and tangent $t=0,1$ to the defect so (\ref{6}) becomes
\begin{equation}\label{7}
\frac{1}{2} \int \text{d}^2x  \;\chi_V \; \partial_t \partial^t z+\frac{1}{2} \int \text{d}^2x  \;\chi_V \; \partial_n \partial^n z.
\end{equation}

To preserve gauge invariance and SUSY in the first term of (\ref{7}), $\chi_V$ should not depend on $x_{0,1}$
so it must be a constant. Note that gauge invariance is preserved up to a total derivative in the $x_{0,1}$-directions which, as we will see later, correspond to the non-zero component of magnetic current. The second term of (\ref{7}), vanishes automatically as we need because there is no dependence on $x_{2,3}$ after dimensional reduction.

As we can see in (\ref{4}) and (\ref{6}), we are dealing with some defect integrations which include D3 worldvolume fields like $D(x_0, x_1, x_4, x_5)$. In such a case, we should set $x_4$ and $x_5$ equal to zero to ensure that the only values of $D$ that contribute to the integrations are those lying on the defect.

After varying with respect to the auxiliary fields to eliminate them, we should add two integrations which contain D-brane fields (e.g., $D$): one is performed on the D3 worldvolume and the other on the defect. So we must rewrite the latter case in this form \footnote{For the delta function in complex coordinates we use the convention $\int d^2z \; \delta^2(z,\bar{z})=1$.}:
\begin{equation}\label{8}
\int \text{d}^2x \; f(x_0,x_1) D(x_0,x_1,0,0)=\int \text{d}^2x \;  \text{d}^2 z_1 \; f(x_0,x_1)\;\delta^2(z_1,\bar{z}_1) D(x_0,x_1,x_4,x_5),
\end{equation}
where $f$ is any function living on the defect namely B, C and $\chi$ or any functions of them. It is clear that the field variations on the D3-brane worldvolume (like $\delta$D), should also be treated in this way.

We mentioned that one can add FI terms to the action without SYM invariance violation but now we demand FI terms to act like a gauge transformation for the theory. This means that adding them to action has no physical effects. For now, It is quite reasonable! It is because, in field theory language, it is convincible to imagine that magnetic charge quantization is a consequence of an extra constraint. In the last section, we will see that the value of $D_V$ on the defect is determined somewhere out of the EOM so it is not a dynamical variable so adding $\int \text{d}^2x \; D_V$ to the $S_{3\text{-}3'}$ means no physics.

Now the bosonic sector of the full action in component form is
\begin{align}\label{9}
&S_{tot}=\frac{1}{g^2_{YM}} \int_{D3} \text{d}^2x \; \text{d}x_4 \text{d}x_5 \Big[-\frac{1}{4} F^{\mu\nu}_V F_{V \mu\nu}+\frac{1}{2} D^2_V\\ \nonumber
&+\frac{1}{2} (Q_3 \partial_{z_1} F_{Q_2}+F_{Q_3} \partial_{z_1} Q_2-Q_2 \partial_{z_1} F_{Q_3}-F_{Q_2} \partial_{z_1} Q_3)+c.c. \\ \nonumber
&+\bar{\partial}_{z_1} A^\mu_V \; \partial_{z_1} A_{V \mu}+\frac{i}{\sqrt{2}}\bar{\partial}_{z_1} A^\mu_V \; \partial_\mu Q_1
-\frac{i}{\sqrt{2}}\partial_{z_1} A^\mu_V \; \partial_\mu \bar{Q}_1 \\ \nonumber
&+\partial^\mu Q_1 \; \partial_\mu \bar{Q}_1-\frac{1}{\sqrt{2}}(Q_1 \bar{\partial}_{z_1} D_V+\bar{Q}_1 \partial_{z_1} D_V)-\frac{1}{2}\bar{\partial}_{z_1} A^\mu_V \; \partial_{z_1} A_{V \mu}+F_{Q_1} \bar{F}_{Q_1} \\ \nonumber
& +\partial^\mu Q_2 \; \partial_\mu \bar{Q}_2+\partial^\mu Q_3 \; \partial_\mu \bar{Q}_3+F_{Q_2} \bar{F}_{Q_2}+F_{Q_3} \bar{F}_{Q_3} \Big] \\ \nonumber
&+\frac{1}{g^2_{YM}} \int _{Def} \text{d}^2x \Bigg[\frac{|B|^2}{4} \Big(2(D_V-D_{V'})+A^2_V+A^2_{V'}-2A_V.A_{V'}\Big)+ \partial^\mu B \; \partial_\mu \bar{B}  \\ \nonumber
&+ \frac{i}{2}(B\partial_\mu \bar{B}-\bar{B} \partial_\mu B)(A^\mu_V-A^\mu_{V'})+F_{B} \bar{F}_{B}+(B \longleftrightarrow C \;\;and \;\; V \longleftrightarrow V')  \\ \nonumber
&+\frac{i}{\sqrt{2}} \Big(BCF_{Q_3}+Q_3(BF_C+CF_B) \Big)+c.c.+ (Q_3 \longleftrightarrow S_3)+c.c.  \\ \nonumber
&+\frac{1}{2} \chi_V D_V+\frac{1}{2} \chi_{V'} D_{V'}\Bigg] \\ \nonumber
&+S_{3'\text{-}3'}.
\end{align}

To eliminate the auxiliary fields in the full action, we should obtain their equations of motion. Without loss of generality, we consider just auxiliary fields on the D3 and obtain
\begin{align}\label{10}
&\bar{F}_B=\frac{iC}{\sqrt{2}}(S_3-Q_3), \\ \nonumber
&\bar{F}_B=\frac{iB}{\sqrt{2}}(S_3-Q_3), \\ \nonumber
&\bar{F}_{Q_2}=\partial_{z_1} Q_3 , \\ \nonumber
&\bar{F}_{Q_3}=-\partial_{z_1} Q_2-\frac{i}{2\sqrt{2}} BC \; \delta^2(z_1,\bar{z}_1), \\ \nonumber
&\bar{F}_{Q_1}=0, \\ \nonumber
&D_V=-\frac{1}{\sqrt{2}}(\bar{\partial}_{z_1} Q_1+\partial_{z_1} \bar{Q}_1)+\frac{1}{4} \; \delta^2(z_1,\bar{z}_1) \big( |C|^2-|B|^2+\chi_V \big),
\end{align} 
where we define $\mathcal{P}\equiv |C|^2-|B|^2$ as the contribution of the charged fields to the D-terms. 

On the D3-brane, the original gauge field strength $F_{ab}$, (with $a,b=0,1,4,5$) obeys the Bianchi identity  $\partial_a \tilde{F}^{ab}=\epsilon^{abcd} \partial_a F_{cd}=0$, everywhere  on  its  worldvolume  and  hence  cannot  carry  a magnetic charge. Instead, in comparison with \cite{Ref6}, we must define the magnetic monopole source term:
\begin{align}\label{11}
&\mathcal{F}_{45} \equiv F_{45}+\frac{1}{4} \; \delta^2(z_1,\bar{z}_1) \mathcal{P}_{FI}, \\ \nonumber
&\mathcal{F}_{ab} \equiv F_{ab} \;\;\; \text{for} \;\;\;a,b\neq 4,5,
\end{align} 
where the appearance of $\mathcal{P}_{FI}$, tells us that moving between two equivalent theories (which differ only by the values of their magnetic charges) by means of a FI-gauge transformation, corresponds to
\begin{equation}\label{12}
\mathcal{P} \longrightarrow \mathcal{P}_{FI}=\mathcal{P}+\chi_V.
\end{equation} 

Now after the elimination of the auxiliary fields by using  (\ref{10}),  it  is  easy  to  show  that  it  is  $\mathcal{F}_{ab}$  which  is  the field strength which appears in the Lagrangian (\ref{9}) in the standard Maxwell form $\mathcal{F}_{ab} \mathcal{F}^{ab}$. The important consequence of definition (\ref{11}) is that the contribution of $|C|^2-|B|^2$ is now absorbed into the kinetic part of the Maxwell-form theory so there is nothing to be concerned about the SUSY breaking by the Fayet-Iliopoulos mechanism \cite{Ref12}.

The electric equation of motion $d*\mathcal{F}=*j_{e,V}$,  can be obtained directly by varying the action with
respect to the gauge field $A_V$, as usual, such that one can show that the only non-zero components
of the electric current are those in the $(01)$ directions.

As a type of “weak test” for our assumption, we can imagine that $\chi_V$ is a definite function (not a field variable) of $x_{0,1}$ and should be determined by an extra constraint, and show that by opposing the FI-gauge invariance to the system, this will become a constant automatically.

It is clear from the first definition in (\ref{11}) that the field strength $\mathcal{F}_{ab}$ is not constrained to obey the Bianchi identity because it is no longer a closed form, so we are dealing with the magnetic equation of motion $d\mathcal{F}=*j_{m,V}$,  which becomes
\begin{equation}\label{13}
\partial_a B^a=j^0_{m,V}=\frac{1}{4} \partial_1 \mathcal{P}_{FI} \; \delta^2(z_1,\bar{z}_1),
\end{equation} 
with $B^a=\tilde{F}^{0a}$, as usual. Notice that the magnetic current $j_{m,V}$ is also tangent to the defect so from the perspective of the D3-brane, one can obtain the magnetic charge which the 1+1-dimensional kink carries by integrating the associated magnetic field $B^a$, over a $\mathbf{S}^2$, surrounding the kink:
\begin{equation}\label{14}
Q_M=\int_{\mathbf{S}^2} \mathcal{F}=\int_{\mathbf{S}^2} B^a ds^a=\frac{1}{2} \Big( \mathcal{P}_{FI}(x^1=+\infty)-\mathcal{P}_{FI}(x^1=-\infty) \Big)
\end{equation} 
where in the second equality we have deformed the $\mathbf{S}^2$, for covering the kink in the $x^1$-direction because of the Dirac delta function.

We need a certain value for $Q_M$ so
\begin{equation}\label{15}
\chi_V (x^1=+\infty)-\chi_V (x^1=-\infty),
\end{equation}
should be canceled out. This leads us to regard $\chi_V$ as a constant as we promised earlier. But FI parameter should be a definite constant, not any constant because as long as it has not been specified, the general action is incomplete and ambiguous and cannot describe a definite physical situation. It is like all definite constants which appear in actions in physics e.g., brane tensions in the DBI action or masses in the standard model action which should be identified by definite values of brane tensions and masses of the elementary particles, respectively. So one should obtain the clear and correct final form of the full theory (action), by determining the FI parameter. For now, we choose the value of $\chi_V$, $8\pi$, by convention to specify the full theory. So equation(\ref{12}) becomes
\begin{equation}\label{16}
\mathcal{P} \sim \mathcal{P}+8\pi.
\end{equation}
This is nothing but the Dirac quantization for magnetic monopoles because it leads to $Q_M=4\pi$, as expected.

The final step is to follow the standard argument in \cite{Ref6} to obtain the quantization condition
(\ref{16}) directly by usual integration of the vector potential like that in the Aharonov-Bohm (AB) effect. $\mathcal{P}$ is a magnetic source for the vector potential because as we will see, the magnetic solution minimizes the action. A suitable gauge transformation leaves a dominant pure gauge term for the vector potential near the defect. In other words,  $\mathcal{P}$ should be in simple pole in the expansion of $A_{Vz_1}$, near the defect, i.e., we should have $A_{Vz_1}=\frac{-i}{8\pi} \frac{\mathcal{P}}{z_1}$. When the point $z_1 = 0$, is extracted and pushed off to infinity such that the remaining space is multiply connected, the holonomy is
\begin{equation}\label{17}
\oint \text{d}z_1 \; A_{Vz_1}+c.c.=\oint (\text{d}x_4 A_{V_4}+\text{d}x_5 A_{V_5})=\frac{1}{2} \mathrm{Re} \mathcal{P}.
\end{equation}

A shift of the holonomy by $4\pi$, gives a physically equivalent configuration, therefore, we again reach (\ref{16}), which means that $\mathcal{P}$ {\it describes the same physical configuration as $\mathcal{P}+8\pi$ does}. In fact, this is why we have regarded the FI terms as a gauge choice. Suppose that adding the FI-term to the system has a kind of physical consequence so (\ref{12}) implies that $\mathcal{P}$ and $\mathcal{P}+8\pi$ are two distinct points in the $\mathbb{R}^1$ space, which is a harsh contradiction with the standard geometrical statement: $\mathcal{P} \sim \mathcal{P}+8 \pi$. So we demanded the consistency and now we see that all things work together successfully. Although the electric charges are also localized in  $z_1 = 0$, it is notable to mention that there is not such a constraint for the electric case.

Note also that one can add another (or more) FI-term(s) to the system again without changing it so $\chi_V$ is a representative of the equivalence class $8\pi n$, where $n$ is an integer.

In the general case, when $\chi_V$  is not equal to $8\pi$, there always exists a scale factor $\lambda$, such that
$\lambda \chi_V=8 \pi$, for determining the quantization condition. So adding $S_{FI}$ to the system, does not change the physics i.e., we have: $\mathcal{P} \sim \mathcal{P}+\lambda \chi_V$ . So we find $Q_M=\frac{\lambda}{2} \chi_V=4 \pi n$. The fact that FI parameter admits only a quantized value has been also predicted in \cite{Ref13} but in a completely different framework.

Let us examine another alternative approach to check this result quickly. We can consider a typical SYM theory which first contains just a vector multiplet which lives on the defect without coupling to any charged matter multiplets and second, the FI-term has been included since the beginning.  It immediately yields the FI parameter quantization
because it appears in the simple pole of the locally pure gauge vector potential (i.e.  we have: $A_{V_z} \propto \frac{\chi_V}{z}$ ) so it can be
excluded easily and leads to $Q_M =0$. Notice that it corresponds to a map from $\mathbf{S}^1$ to $U(1)$, which is specified by $n=0$, i.e., $n=0$, represents the homotopy class to which this bundle belongs.

Finally, it is obvious that the whole argument is also true for the D3'-brane.

\section{A short action-free analysis}

Aside from our action analysis, one can also imagine that we are dealing with the SYM version of the Wilson loop which can be regarded as a physical quantity, namely
\begin{equation}\label{18}
\int _{D \text{-}term \; over \; defect} V \; \text{d}^2x,
\end{equation}
instead of its contribution to the action.

The action is intrinsically irregular because of the appearance of a quadratic delta divergence. But the electric-magnetic duality tempts us to try to regularize and minimize the action simultaneously. Fortunately, this is possible. With a suitable gauge choice for $A_\mu$, near the defect, we can eliminate the divergence so $D_V$ is fixed on the defect which guarantees that this choice minimizes the action. This is why $S_{FI}$ means no physics. It implies that $D_V$ is a definite function (not a field variable) over the region of the singularity i.e., we have $\delta_{virtual}D|_{def}=0$. In other words, $D_V|_{def}$ determines the zero point of energy and as we will explain, it is governed by a geometrical constraint rather than a dynamical one. By ``singularity'' we mean a space-time singularity i.e., removing it from $\mathbb{R}^3$ yields a manifold with the same homotopy type as $S^2$, so it should not be confused with the moduli space singularities may apear in some other theories. $Q_1$ is not an analytic function in this region.

In an ordinary case, we should extract the singularity during the integration over the Wilson loop while in the SYM version, the integration is performed over the singular region. Look how SUSY allows us to probe around singularities because of $\int \text{d}^4 \theta$.

But how can we show that (\ref{18}), which is both SUSY and gauge invariance, can be called $Q_M$? On the D3 worldvolume and near the defect, a suitable gauge transformation leads to a dominant pure gauge term which can eliminate the quadratic delta divergence. The general form of $Q_1$, which leads to a non-vanishing finite contribution to $D_V$ on the defect, is
\begin{equation}\label{19}
Q_1(x^a) \approx \frac{1}{8 \pi \sqrt{2}} \frac {\mathcal{P}}{z_1}- \sqrt{2} \; \zeta(x^0,x^1) \bar{z}_1,
\end{equation}
i.e.,  $\mathcal{P}$ should be in the simple pole to cancel the delta function divergence and the effect of this special gauge transformation on the remainder is negligible near the defect.  All other positive
powers of $z_1$ and $\bar{z}_1$  or mix of them have no contribution to $Q_1$ (and also $D_V$) over the
defect. For the $\zeta$-term which has no contribution to $Q_1$ (but for $D_V$ it has) near the defect, we should discuss in more detail.

Definition (\ref{19}) leads to (\ref{16}) again so, as we mentioned earlier, $D_V (x_0,x_1,0,0)$ (which is equal to $2 \mathrm{Re} \zeta$), is a definite function. It allows us to define the magnetic flux $Q_M$ as follows:
\begin{equation}\label{20}
Q_M \equiv \int_{singular \; region} \text{d}^2x  \; \text{Re}\zeta(x^0,x^1).
\end{equation}

We can keep on moving forward by using a quantum concept: in the functional integral, we have a factor $e^{\frac{iS}{2}}$, such that two actions are equivalent if they differ by $8\pi n$. So once the number of the magnetic charge quanta $n$ (or the winding number) is specified, we end with the equation which governs the magnetic flux distribution \footnote{It should not be confused with the magnetic charge density $j^0_{m,V}$.}:
\begin{equation}\label{21}
\int_{D\text{-}term \; over \; singular \; region} V \; \text{d}^2x=\frac{1}{2} \int_{singular \; region} \text{d}^2x \;  D_V=Q_M=4 \pi n.
\end{equation}

For the regions with $\mathrm{dim}_{def}>1+1$, we should insert a suitable $M_{Planck}$ in the RHS of (\ref{21}) to make it dimensionless so the case of $\mathrm{dim}_{def}=1+1$ seems to be very nice for the phenomenological purposes. If $\mathrm{dim}_{space\text{-}time}=l$ and $\mathrm{dim}_{def}=m$, this interesting region is characterized by $m=\frac{l}{2}$.

One of the simplest choices for $\zeta(x^0,x^1)$ which represents the class of magnetic monopole quanta,
 is $\delta^2(x^0,x^1)$. The allowed choices are some special sort of functions which are classified
by an integer $n$ which corresponds to the homotopy class of the non-trivial $U_V(1)$ bundle. The volume surrounded 
between these functions and coordinate axes, must be equal to this integer. It is obvious that first, this
 structure has group properties with respect to addition (in fact, this is isomorphic to $\frac{\mathbb{Z}}{[0]}$) and second, a change of coordinates does not affect this integer. We should keep in mind that constant coefficients should be dropped before the evaluation of an integral, e.g., $3 \zeta(x^0,x^1)$ is the same as $\zeta(x^0,x^1)$.

The definiteness is an intrinsic characteristic for the system which plays a crucial role in the analysis so if one can construct a situation which preserves gauge invariance of (\ref{20}), then he or she can use this picture.

Our action analysis is the only alternative way to reach the magnetic charge quantization again, from the perspective of action and without paying any expenses. Moreover, by remembering Noether's theorem, we know that electric and magnetic charges are associated to gauge invariance so one may think that {\it the quantization} of magnetic charges maybe is a consequence of an extra unusual symmetry which should be related to SUSY and gauge invariance of the theory in some special manner. Note that the magnetic current conservation is manifested because we have: $d^2 \mathcal{F}=d*j_M=0$. In the case of doubt, the reader can also check this conservation directly by using the explicit form of the components of $j_M$ and show that $\frac{d}{dt} Q_M=0$.

Ultimately, the general result is not surprising but expectable because when holonomy is non-trivial and the gauge group of a theory breaks to U(1)'s, one can obtain magnetic charge quantization by integration of solutions of BPS equations over some boundary regions. Since these BPS solutions respect SUSY, one can guess there should be a different way to reach magnetic charge quantization directly by SYM invariance arguments.

Note again that in this paper we did not talk about any magnetic solutions which, as  explained in \cite{Ref6} for the D3-D3' case, can be obtained by considering a non-canonical kinetic term and solving BPS equations. We discussed  about  quantization  of  magnetic  charges  which these solutions carry as a topological property  and  showed  that  there  is  another  way  to  reach it.

\section{Conclusion} \label{conclu}

AB effect for the magnetic flux quantization is based on quantum mechanics. In 1975, Wu and Yang derived the same result by a stronger systematics based on the notion of fiber bundles and showed that the magnetic charge quantization is best described in a topological manner i.e., the Dirac monopole can be considered as a $U(1)$ bundle over a $S^2$. In this approach, the wave function and its single valuedness which were playing the main role in AB effect, is just the section of the corresponding line bundle.

In this paper, we have mentioned that there is an alternative approach to obtain the magnetic charge quantization condition for the D3-D3' set-up. This prescription is based on SUSY and gauge invariance properties of the system such that they originate the Fayet-Iliopoulos gauge invariance of the configuration.
 
This picture also shows that the FI parameter is nothing but the magnetic charge itself (up to a constant) so it should be a quantized value. So one can imagine that we are dealing with SYM statement of Stokes' theorem. In general, at any arbitrary time, we can drop the time integration of the measure of (\ref{21}), so we have:
\begin{equation}\label{22}
2\pi c_1(F)=\int_{\mathbf{S}^2} F  \; \longleftrightarrow \int_{D\text{-}term \; over \; singular \; region} V \; \text{d}^dx,
\end{equation}
where $c_1(F)$ is the first Chern class and the integration of $V$ is  preformed  over  the  spatially  extended  singular region. It is obvious that if the configuration does not contain a topological defect, then both sides of (\ref{22})  become zero and there  exists no  magnetic charge  as expected.

It seems hard to generalize this procedure to the non-Abelian case because, as we mentioned earlier, adding FI terms is allowed only when the gauge group is $U(1) \times \cdots \times U(1)$. In the non-Abelian case, contents mixture is such that one cannot find any quantization condition for the system. But in the Coulomb branch of a gauge theory where $U(N)$, breaks to $U(1) \times \cdots \times U(1)$, one can try to examine how it works.

Another interesting feature is that the magnetic charge quantization does not depend on whether we are in the canonical or non-canonical description of the system, as expected. It is because in latter case we must add FI-terms to the K\"ahler potential $K(Be^{(V-V')},\bar{B},Ce^{(V'-V)},\bar{C})$ for kinetic terms as before.  After doing that one can show that the effect of FI terms equals to the canonical description and leads to (\ref{16}) again for a new $\mathcal{P}$.



\end{document}